\documentclass[aip,reprint]{revtex4-1}

\usepackage{graphicx}
\usepackage{color}

\draft 

\begin{document}

\title{Sample size dependence of tagged molecule dynamics in steady-state networks with bimolecular reactions: Cycle times of a light-driven pump}

\author{Daniele Asnicar}
\affiliation{Department of Chemical Sciences, University of Padova, via Marzolo 1, I-35131 Padova, Italy}
\author{Emanuele Penocchio}
\affiliation{Department of Physics and Materials Science, University of Luxembourg, avenue de la 
Fa\"{ı}encerie, Luxembourg City, L-1511, G.D. Luxembourg}
\author{Diego Frezzato}
\email{diego.frezzato@unipd.it}
\affiliation{Department of Chemical Sciences, University of Padova, via Marzolo 1, I-35131 Padova, Italy}


\begin{abstract}
Steady-state reaction networks are here inspected from the viewpoint of individual tagged molecules jumping among their chemical states upon occurrence of the reactive events. Such an agent-based viewpoint is useful for selectively characterizing the behaviour of functional molecules, especially in the presence of bimolecular processes. We present the tools for simulating the jump dynamics, both in the macroscopic limit and in the small-volume sample where the numbers of reactive molecules are of the order of few units with an inherently stochastic kinetics. The focus is on how an ideal spatial `compartmentalization' may affect the dynamical features of the tagged molecule. The general approach is applied to a synthetic light-driven supramolecular pump constituted by ring-like and axle-like molecules that dynamically assemble and disassemble originating an average ring-through-axle directed motion under constant irradiation. In such an example, the dynamical feature of interest is the completion time of direct/inverse cycles of tagged rings and axles.
We find a surprisingly strong robustness of the average cycle times with respect to the system's size. This is explained with the presence of rate-determining unimolecular processes, which may therefore play a crucial role in stabilizing the behavior of small chemical systems against strong fluctuations in the number of molecules.
\end{abstract}

\maketitle
\section{Introduction}

Several key functions in biochemical contexts are regulated by networks of chemical reactions taking place in a fluid and thermostated environment under out-of-equilibrium steady state conditions. A privileged viewpoint is that of an individual tagged molecule, or even a tagged molecular fragment (a moiety), that changes its chemical state when involved in a reaction. Following the fate of a tagged molecule offers an insight on subtle features that would be otherwise shadowed if looking at the global evolution of the whole reactive system\cite{Ninio, skodje1, skodje2, skodje3, sabatino2019a, frezzato2021}. In particular, one can access specific statistical descriptors like the distribution of first occurrence times of certain reactive events involving the individual molecule, or the distribution of completion times of cyclical processes, and so on. Think, for example, to the statistics of the turnover time for an individual enzyme molecule involved in a catalytic scheme\cite{Xie2005,moffit2014}, or to the statistical kinetics of processive enzymes\cite{Block95} and molecular motors\cite{Fisher07}. Such information is much more detailed than the mere average rate of the products formation at the steady state. Moreover, by adopting the agent-based viewpoint of a tagged molecule, one focuses precisely on the behaviour of the functional part of the whole machinery.

The specific `function' of a tagged molecule can be characterized by a dynamical output like, for instance, the cycle time of an enzyme molecule, or the quantitative descriptor of a much more articulated event (the concept will be further elaborated later).
By viewing each chemical state as a `site', the tagged molecule follows its path among the available sites. Because of the randomness on the sequence of reactive events involving the tagged molecule, and on the time at which each reaction will take place, such a path is a stochastic wandering under steady state conditions. Hence, any dynamical output can be characterized only in statistical terms (distribution of the outcomes, average values and standard deviations, correlation functions, etc.). 

An issue to be explored is the sensitivity of tagged molecules dynamical features with respect to the size of the sample. In fact, one can pass from the macroscopic limit in which the huge numbers of reactive molecules guarantee that deterministic kinetic laws apply \cite{Laidler}, to the case of volumes so small that the numbers of molecules are in the order of units or tens and the kinetics is inherently stochastic\cite{gillespie2007, gillespie2013perspective, grima2017}. The question to pose is: {\em Under steady conditions, how does a dynamical output of a tagged molecule change in passing from the macroscopic limit to the small-volume case?} In particular: {\em Is there a marked dependence on the size of the sample?} Of course, for exploring the size dependence in a meaningful way one has fix the relative abundance of the species involved. This means to imagine a sort of `compartmentalization' in which, under steady conditions, a small portion of the well-mixed solution is suddenly isolated from the rest of the macroscopic sample, and let free to evolve. Thus, the above questions can be reformulated as: {\em Does the compartmentalization bring some effect on the dynamical output of a tagged molecule under steady conditions?} This is the problem faced in the present work.

Compartmentalization effects on the rates of biochemical reactions are currently under intense experimental study, for instance using `coacervates' (artificial cell-like environments constituted by water-in-oil microdroplets) as microreactors\cite{coac1, coac2}. This motivates the development of theoretical/computational tools for inspecting the transition from macro- to microscale, in particular by adopting the highly informative individual-molecule perspective. However, we must stress from the beginning that one basic assumption here is that all {\em intrinsic} kinetic parameters (i.e., the parameters associated with reaction coordinates and energetics at the molecular scale) remain unaltered by the change of sample's size. In real situations, the compartmentalization might subtly alter also the kinetic parameters, for instance by inducing a stabilization of transition states as suggested for some reactions in coacervates\cite{coac1, coac2} in which the change of local polarity could be very relevant. Such a kind of additional case-dependent effects are ignored in the present study.

A crucial aspect is the fact that a volume dependence can manifest itself only in the presence of bimolecular reactions. This is because the so-called reaction `propensities' depend on the sample volume only for nonlinear processes, as it will be made explicit later. The choice of a relevant case study is thus oriented towards reaction networks in which the kinetics of the tagged molecule involves bimolecular processes such as self-assembly. Apart from being widespread in natural biochemical networks, self-assembly is a key feature in the broad fields of systems chemistry and artificial supramolecular machines, in which specific tasks are realized thanks to a proper design of the molecular players and a suitable choice of the operative conditions\cite{Credi1,Credi2,Credi3,Credi4,Leigh1,Leigh2,Astumian1}. 

The system explored here is the first prototype of synthetic light-driven pump constituted by ring-like and axle-like molecules. When put together in solution, these species dynamically assemble and disassemble originating, under constant irradiation in the UV-Vis range, an average directionality in the dynamics of a tagged ring through the axles, or of a tagged axle through the rings. The directionality arises because the irradiation breaks the detailed balance by keeping the system in out-of-equilibrium steady conditions. 
Such a system has been synthetized and characterized by Credi and co-workers\cite{ragazzon2015}, 
and recently subjected to chemical modifications aimed at improving the efficiency and the possibility of chemical functionalization of the molecular machinery\cite{canton2021second}. 
Among various possibilities, the dynamical output considered here is the cycle time of a tagged ring or of a tagged axle in one direction (direct cycles) and in the opposite one (inverse cycles), as it will be defined later. The statistical analysis of the cycle times was done in the macroscopic limit\cite{sabatino2019b}. 
Here, we extend the analysis by exploring also the small-volume context.
We notice that size effects in the functioning of synthetic molecular motors have not been explored in previous computational studies\cite{Feringa09, Gingrich21}, as the latter focused on the much more common case of unimolecular motors.

Our purpose is twofold. First, we wish to present the general methodology for the tagged-molecule analysis. The matter has been partially illustrated elsewhere\cite{sabatino2019a, frezzato2021}, but a self-contained presentation proves useful to cover both the macroscopic and the small-volume instances. With the simulation tools at hand, we then focus on the ring-axle case. By means of a phenomenological inspection based on simulations, we aim at understanding if the compartmentalization might add a dimension to the control of the dynamical behaviour of the individual molecules.  

The paper is arranged as follows. First, the tagged-molecule viewpoint is presented in all generality in Sec. \ref{sec_tagged}, where we specify what `steady conditions' means in the macroscopic and small-volume limit, introduce the jump dynamics of a tagged molecule among its available chemical states, and separately characterize such dynamics in the macroscopic and small-volume situations. This will provide the general methodological framework. The specific ring-axle case is introduced in Sec. \ref{sec_case_model}, framed under the tagged-molecule viewpoint in Sec. \ref{sec_stat}, and numerically inspected and commented in Sec. \ref{sec_numerical}. Section \ref{sec_results} is devoted to concluding remarks.

\section{Reaction networks from the viewpoint of a tagged molecule}\label{sec_tagged}

\subsection{Steady conditions}

Let us first specify the physical conditions under which we characterize a reaction network from the viewpoint of tagged molecules. We refer to the ideal situation of a fluid, thermostated and spatially homogeneous reaction environment of constant volume $V$ in which a pool of reactions, involving $N$ chemical species, take place at the steady state (superscript `ss'). Under steady conditions, both in the macroscopic and in the small-volume contexts, all rate constants entering the macroscopic kinetic laws must be time-independent. In addition, sink processes must be absent (otherwise some species would disappear) or must be compensated by external restoring actions.

In the macroscopic limit, the steady state of the whole system is fully specified by the volumetric concentrations of the chemical species, which we collect in the array
\begin{equation}
{\bf x}^{\rm ss} = \left( x_1^{\rm ss}, x_2^{\rm ss}, \cdots, x_N^{\rm ss} \right)    
\end{equation}
Such a state can be  numerically determined by finding the stationary point of the ordinary differential equations of the macroscopic kinetics, for instance by making a time propagation up to reach the long-time limit.

In the small-volume context, the system continues to fluctuate even under the steady condition. At a time $t$, the system's state is specified by the array 
\begin{equation}
{\bf n}(t) = \left[ n_1(t), n_2(t), \cdots, n_N(t) \right]   
\end{equation}
where $n_i(t)$ is the number of molecules of the $i$-th species. 
However, after a long time from a given initial condition ${\bf n}_0 = {\bf n}(0)$, the fluctuations in the space of the copy numbers do settle on a time-independent distribution. In terms of conditional probability, we can write
\begin{equation}
\lim_{t \to \infty} p({\bf n}, t | {\bf n}_0) = p^{\rm ss}({\bf n})    
\end{equation}
for any ${\bf n}_0$ (we assume that the stationary distribution is unique).
This is what we mean by `steady conditions' in the small-volume context.

While in the macroscopic context the data acquisition for any statistical analysis can start at an arbitrary time, in the small-volume context one has to neglect the initial transient regime during which the copy numbers of each species are correlated with the initial condition ${\bf n}_0$.

\subsection{Jump dynamics among sites}\label{sec_jumps}

We now introduce the tagged-molecule viewpoint in all generality. Assume to be under the steady conditions specified above. Now suppose to be able to follow the evolution of a tagged molecule among the many of the same kind that are present in the reaction environment. 

An example proves useful. We might think to follow a tagged enzyme molecule (species E) in a solution where a substrate (S) is converted into a product (P) according to the basic Michaelis-Menten catalysis 
${\rm E} + {\rm S} \leftrightarrow {\rm ES} \leftrightarrow {\rm E} + {\rm P}$
(in all generality we assume that the product formation is reversible; if the product formation were an irreversible sink process, the substrate should be restored by an external chemostating action to maintain the steady state). We would see the enzyme molecule jumping between two sites: one corresponding to the free molecule (site 1), the other to the molecule bound in the ES complex (site 2). A jump from site 1 to site 2 occurs when the tagged molecule is involved in the reaction ${\rm E} + {\rm S} \to {\rm ES}$ or in the reaction ${\rm E} + {\rm P} \to {\rm ES}$. A jump from site 2 to site 1 occurs when the complex ES hosting that molecule is involved either in the reaction ${\rm ES} \to {\rm E} + {\rm S}$ or in the reaction ${\rm ES} \to {\rm E} + {\rm P}$.   

The above example serves to give the idea that from a pool of reactions occurring in parallel one can extract a jump process for the tagged molecule among its hosting sites. One could go even further by considering, instead of a whole molecule, a chemical moiety (e.g., a tagged atom or a tagged functional group) that jumps from site to site\cite{sabatino2019a, frezzato2021}. For simplicity, we proceed by referring to tagged molecules, but it is meant that the whole approach can be easily extended to tagged moieties. 

In all generality, one ends up with a finite set of sites. 
If  $s$ and $s'$ are two sites, the connection between them is established by the presence of a chemical reaction $m$ bringing the molecule from site $s$ to site $s'$ (and vice-versa in case of reversible reactions). The path of a tagged molecule is thus a sequence of the kind
\begin{eqnarray}\label{eq_seq}
\cdots s_{(t_1, t_2]} \stackrel{m(t_2)}{\to} s_{(t_2, t_3]} \stackrel{m(t_3)}{\to} \cdots
s_{(t_n, t_{n+1}]} \stackrel{m(t_{n+1})}{\to} \cdots
\end{eqnarray}
where the notation $s_{(t_a, t_b]}$ stands for site occupied in the time interval $t_a < t \leq t_b$, and $m(t)$ indicates the reaction that occurs at the time $t$. It is meant that the reactive events are instantaneous.
This requires to assume that the sites correspond to pools of microstates pertaining to separated wells of configurational free energy, and that there is a marked timescale separation between slow site-to-site transitions and much faster intra-well relaxation processes. In such a limit, the site-to-site transitions occur rarely, and the jumps have ideally `no duration'.
Note that there might be several reaction channels that allow the jump from one site to another; for instance, a certain process could occur thermally or via a photochemical pathway. Depending on the specific purpose, the multiple reaction channels can be kept differentiated one from the other, or merged to some extent. In case of merging, the specifications $m(t)$ on the top of the arrows in Eq. \ref{eq_seq} should be replaced by a group label (or removed in case of complete merging).

Given a long sequence like Eq. \ref{eq_seq} under steady conditions, one can characterize, in statistical terms, any dynamical output in which the tagged molecule is involved as an `agent'. The expression `dynamical output' stands for any quantitative output, or even for any event expressed by a verbose statement, that can be determined/assessed from the sequence. For instance, turning back to the Michaelis-Menten kinetics from the viewpoint of a tagged enzyme molecule, the dynamical output could be the next formation of a product molecule, or could be a more elaborated statement of the kind `After the formation of a product molecule, the enzyme must jump forth and back from E to ES for three times, then the next formed ES has to give P, and all of this must take place in a time not longer than $t^\ast$'. One might ask which is the probability of observing such a complex event within a given time window of inspection.
While it would be very difficult, or even impossible, to give an answer to similar questions by trying to set the problem on pure probabilistic basis, i.e. by reasoning in terms of the associated master equations incorporating all the clauses, it proves simple to attack the problem by simulating sequences like Eq. \ref{eq_seq} and then making the post-production statistical analysis. 

The nature of the jump dynamics among the available sites depends on the size of sample. In the macroscopic limit, the dynamics of a tagged molecule reduces to a memoryless Markov jump process with fixed jump rate constants, as detailed in Sec. \ref{sec_macro}. In the small-volume context, treated in Sec. \ref{sec_micro}, the Markov property applies only to the evolution of the whole system in the space of copy numbers but not to the dynamics of a single tagged molecule, unless in presence of only unimolecular reactions (see more comments in the following). Of course, the small-volume situation is more general and formally includes the macroscopic case in the limit of large numbers of molecules.

\subsection{Macroscopic sample}\label{sec_macro}

Let us consider a tagged molecule in the site $s$ corresponding to the species $j$, or being a part of the species $j$ (e.g., the species $j$ may be a complex, or a polymer, etc., in which the tagged molecule is one the components). Let $s'$ be a site connected with $s$. The jump rate constant from $s$ to $s'$, due to the reaction $m$ involving the species $j$, is expressed by the following ratio: 
\begin{equation}\label{jump_rate}
c_{s \stackrel{m}{\to} s'} = \frac{\nu_{R_j}^{(m)} \,  r_m({\bf x}^{\rm ss})}{x_j^{\rm ss}}    
\end{equation}
where ${\nu_{R_j}}^{(m)}$ is the stoichiometric coefficient of the species $j$ as reactant in the reaction $m$, and $r_m({\bf x}^{\rm ss})$ is the rate of such a reaction at the steady state (expressed according to the mass action law)\cite{sabatino2019a}. In fact, the numerator gives the average number of molecules of species $j$ that react in the time unit per unit of volume, while the denominator gives the average number of available molecules per unit of volume. Thus the ratio in Eq. \ref{jump_rate}, when multiplied by a small time interval $\delta t$, gives the probability that in such time interval the specific molecule of species $j$ will be involved in a reaction $m$, and hence the tagged molecule be transferred from $s$ to $s'$. 

As example, suppose that the tagged molecule is a molecule of a species A which can undergo the dimerization ${\rm A} + {\rm A} \stackrel{k_{\rm dim}}\rightarrow {\rm A}_2$. Let site 1 be the molecule in the monomer state A, while site 2 is the molecule in the dimer ${\rm A}_2$. Given that the steady-state rate of the reaction is $k_{\rm dim} [{\rm A}]_{\rm ss}^2$, from Eq. \ref{jump_rate} we get $c_{1 \stackrel{\rm dim}{\to} 2} = 2 k_{\rm dim} [{\rm A}]_{\rm ss}$. Now consider the reverse unimolecular reaction of dissociation, ${\rm A}_2 \stackrel{k_{\rm diss}}\rightarrow {\rm A} + {\rm A}$. The species $j$ here corresponds to the dimer and the tagged molecule is part of it. According to Eq. \ref{jump_rate} we have that $c_{2 \stackrel{\rm diss}{\to} 1} = k_{\rm diss}$.
With this simple rule, from any reactions network at the steady state we can specify the associated jump process between sites for a given tagged molecule. 

We stress that for site-to-site jumps due to unimolecular reactions, the jump rate constants $c_{s \stackrel{m}{\to} s'}$ correspond to the reaction rates. For jumps due to reactions of higher molecularity, the  jump rate constants bear a dependence on the composition of the reactive mixture at the steady state through Eq. \ref{jump_rate}. Such a composition dependence is a crucial fact that can be exploited, in principle, to regulate the dynamical output of a tagged molecule when multimolecular reactions are present. 

Once the number of sites and the jump rate constants are fixed, the stochastic path among the sites can be generated by means of the standard Gillespie's stochastic simulation algorithm (SSA). Suppose that the tagged molecule currently is in the site $s$. In the SSA terminology, the propensity of jumping into a site $s'$ thanks to the reaction $m$ corresponds to $c_{s \stackrel{m}{\to} s'}$ (in fact, $c_{s \stackrel{m}{\to} s'} \, \delta t$ gives the the probability that such a jump takes place in the next time interval $\delta t$). The total propensity of jumping out from the actual site $s$ is then 
\begin{equation}
a_{\rm tot}(s) = \sum_{s'} \sum_m c_{s \stackrel{m}{\to} s'}    
\end{equation}
The time $\tau_{\rm jump}$ after which such a jump occurs is a random variable with distribution\cite{gillespie1977exact} 
\begin{equation}
\rho(\tau_{\rm jump} | s) = a_{\rm tot}(s) \, e^{- a_{\rm tot}(s) \, \tau_{\rm jump}} 
\end{equation}
For generating a path among the sites, we can simply iterate the following steps: (1) randomly generate $\tau_{\rm jump}$ from $\rho(\tau_{\rm jump} | s)$ (this is conveniently done by drawing a number $u$ at random from the uniform distribution between 0 and 1, and then computing $\tau_{\rm jump} = {a_{\rm tot}(s)}^{-1} \, 
\ln \left[ 1/(1-u) \right]$); (2) randomly pick one among all possible channels $s \stackrel{m}{\to} s'$ (just imagine to divide the unit segment into intervals of width
$c_{s \stackrel{m}{\to} s'} / a_{\rm tot}(s)$, draw at random a number from the uniform distribution between 0 and 1 and look at which intervals it falls in; in one stroke, this selects both the arrival site $s'$ and the reaction $m$);
(3) update time (leap of $\tau_{\rm jump}$) and site (from $s$ to $s'$).

\subsection{Small-volume sample}\label{sec_micro}

In a volume sufficiently small that the numbers of reactive molecules are of the order of few units or tens, the evolution of the array ${\bf n}(t)$ is ruled by the laws of stochastic kinetics\cite{gillespie2007, gillespie2013perspective, grima2017}. 
The basic assumptions are that the spatial distribution of each molecule is uniform, the reactive events are treatable as instantaneous jumps, and only one event at a time can occur.
The key quantities are the so-called `propensity functions' of each reaction. For the $m$-th reaction, let $a_m({\bf n})$ be the associated propensity function which depends on the actual state of the global system in the space of the copy numbers.
The propensities, when multiplied by a small $\delta t$, express the probability that a certain reactive event does occurs in the sample in the next time interval $\delta t$. In practice, only unimolecular and bimolecular reactions are physically relevant in real kinetic schemes; the possible cases and the corresponding propensity functions are 
\begin{eqnarray}
&&{\rm A} \stackrel{k_{\rm uni}}{\rightarrow} {\rm P} \; : \;
a_{\rm uni}({\bf n}) = f_{\rm uni} \, n_{\rm A} \cr
&&{\rm A} + {\rm A} \stackrel{k_{\rm bim, 1}}{\to} {\rm P} \; : \;
a_{\rm bim, 1}({\bf n}) = f_{\rm bim, 1} \, n_{\rm A} (n_{\rm A}-1) \cr
&&{\rm A} + {\rm B} \stackrel{k_{\rm bim, 2}}{\to} {\rm P} \; : \;
a_{\rm bim, 2}({\bf n}) = f_{\rm bim, 2} \, n_{\rm A} n_{\rm B}
\end{eqnarray}
where `P' stands for products. In these expressions, the factors $f$, having physical dimension of inverse-of-time, are given by
\begin{eqnarray}\label{fprop}
f_{\rm uni} = k_{\rm uni} \;\; , \;\,  f_{\rm bim, 1} = \frac{k_{\rm bim, 1}}{V} 
\;\; , \;\, f_{\rm bim, 2} = \frac{k_{\rm bim, 2}}{V} 
\end{eqnarray}
where $k_{\rm uni}$, $k_{\rm bim, 1}$ and $k_{\rm bim, 2}$ are the common rate constants of the macroscopic reactions and $V$ is the volume of the sample. 
\footnote{If the bimolecular rate constants are expressed in $\rm M^{-1} s^{-1}$, the volume (in liters) at the denominator must be multiplied by the Avogadro number.}
Such form and parametrization of the propensity functions ensure the correct matching between stochastic kinetics in the small-volume (small copy numbers) context and deterministic mass-action-law based kinetics in the macroscopic limit.

The stochastic path ${\bf n}(t)$ in the space of the copy numbers can be generated by means of Gillespies's algorithm in its original form\cite{gillespie1977exact, gillespie2007, gillespie2013perspective}. 
In short, let $a_{\rm tot}({\bf n}) = \sum_m a_m({\bf n})$ be the total propensity of leaving the current state $\bf n$. The path is generated by iterating the following steps: (1) randomly generate the time $\tau_{\rm react}$ of the next reactive event from the distribution 
$\rho(\tau_{\rm react} | {\bf n}) = a_{\rm tot}({\bf n}) \, e^{- a_{\rm tot}({\bf n}) \, 
\tau_{\rm react}}$; (2) randomly pick the reaction $m$ (each reaction has relative propensity 
$a_m({\bf n})/a_{\rm tot}({\bf n})$ to occur); (3) update the state by taking into account the variation of each copy number due to the firing of the reaction $m$. On the top of such collective evolution, the path of the tagged molecule has to be generated by giving the chance to such a molecule, among the many others of the same type currently present in the sample, of being involved in the next reaction event. It suffices to apply the following rules:
\begin{enumerate}
\item
If the tagged molecule does not enter as reactant in the next reaction, then the state of the molecule remains unaltered.
\item
If the tagged molecule enters as reactant in the next reaction, then the state of the molecule {\em may} change with probability $1/N^\ast$ where $N^\ast$ is the current number molecules of the same type.
\item
Make the time advancement of $\tau_{\rm react}$ and update the state of the tagged molecule.
\end{enumerate}

As anticipated in Sec. \ref{sec_jumps}, in the small-volume situation the jump dynamics of the tagged molecule is not, generally, a Markov process. In fact, while the collective dynamics ${\bf n}(t)$ is a Markov process, the focus on the individual molecule (steps 1-3 above) makes the propensities of the site-to-site jumps possibly bearing a dependence on the actual copy numbers. This implies that the site location alone cannot fully specify the dynamics of the tagged molecule (and so, in the jargon,  
`memory effects' would be introduced). The exception is the case in which all reactions are unimolecular. In such a case, one sees that the site-to-site jumps are associated with jump rate constants coinciding with the reaction rate constants, and the jump dynamics is a Markov process.   

Summarizing, the presence of bimolecular reactions makes that the site-to-site jump dynamics of a tagged molecule in the small-volume context is not a Markov process and, more importantly for our discussion, that a dynamical output of interest depends {\em a priori} on the volume of the sample.

\section{Case study: A light-driven molecular pump}\label{sec_case_model}

\begin{figure*}[ht]
    \centering
    \includegraphics[width=0.8\textwidth]{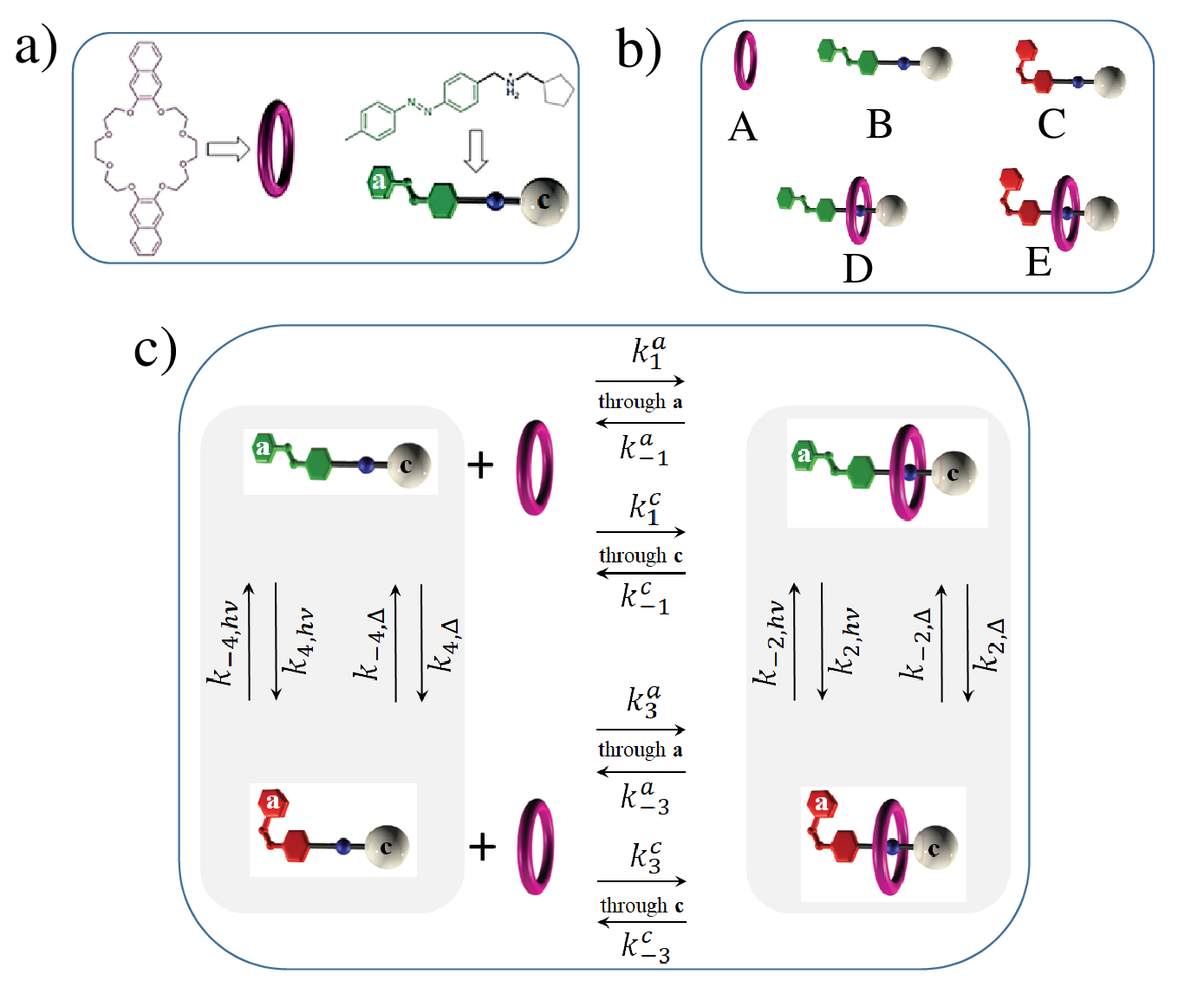}
    \caption{Schematic representation of the involved molecules (panel a), labeling of the species (panel b), and the network of reactions (panel c).}
    \label{fig1}
\end{figure*}

The system here investigated consists of a well-mixed and thermostated solution of rings and axles (see Fig. \ref{fig1}a) which can self-assemble forming a  pseudo-rotaxane. Throughout, ``ring'' stands for 2,3-dinaphtho [24] crown-8 ether, while ``axle'' refers to a molecule comprising  a photoswitchable azobenzene unit at one side (the end {\bf a}), a central ammonium recognition site for the macrocycle (the unit {\bf b}), and a passive methylcyclopentyl pseudo-stopper at the other side (the end {\bf c}) \cite{ragazzon2015,baroncini2012photoactivated}.
The axle has a steric footprint only at one side and so the structure can assemble and disassemble.  The azobenzene can be in the conformation {\em E} ({\em trans}) or {\em Z} ({\em cis}). Altogether, five chemical species are involved, which we label as indicated in Fig. \ref{fig1}b. 

When the azobenzene is in the conformation {\em E}, the ring enters and exits preferably from the azobenzene side, while the use of the methylcyclopentyl side is prevented by steric hindrance. On the other hand, if the conformational transition from {\em E} to {\em Z} takes place when the ring is bound to the axle, the exit from the methylcyclopentyl side becomes the less impeded way. The coupling between self-assembly and photo-induced {\em E}-{\em Z} transitions makes that, on average, the motion of  rings through axles, and of axles through rings, occurs preferably in one direction.
In this perspective, each single ring can be viewed as a molecular machine whose function is being crossed by axles preferably oriented in one way; similarly, each axle is like a machine whose function is being crossed by rings preferably in one direction. Such a desired directionality is achieved only under constant irradiation (in the UV-Vis range) which keeps the reaction mixture in a nonequilibrium photostationary state. On the contrary, at thermal equilibrium no directed motion can be observed.  

The network of reactions involving the five species is shown in Fig. \ref{fig1}c. Table \ref{table1} reports the values of the rate constants, the same 
\footnote{
Unfortunately, the numerical value of the rate constant $k_{-4,\Delta}$ was incorrectly typed as $0.14 \times 10^{-6} \, {\rm s^{-1}}$ in Table 1 of ref. \cite{sabatino2019b}. The correct value is the one reported here, which has been employed for the numerical simulations in both the works. The outcomes presented in ref. \cite{sabatino2019b} are essentially unaffected when using the typed value, 
because the associated jump is regulated by the much larger $k_{-4, h\nu}$, thus preventing any reproducibility issue.
}
employed in ref. \cite{sabatino2019b}. Some reactions are pure thermal processes (rate constants with the subscript `$\Delta$') while others are photo-activated processes (subscript `$h \nu$'); irradiation wavelength and photon flux intensity are the same as for the experimental conditions of ref. \cite{ragazzon2015}.
Note that we make a distinction between entrance/exit of the ring through the azobenzene side of the axle (side \textbf{a}) and entrance/exit through the methylcyclopentyl side (side \textbf{c}); the corresponding bimolecular reactions have different rate constants (subscripts `a' and `c' refer to the specific axle's side).

\begin{table}[h]
    \caption{The employed rate constants. The values for the photochemical processes (subscript ‘$h\nu$’) correspond to the experimental conditions of ref. \cite{ragazzon2015}: irradiation wavelength of 365 nm and photon flux of $1.67\times 10^{-9}$ Einstein s$^{-1}$.}
    \label{table1}
    \begin{ruledtabular}
 \begin{tabular}{ccc} 
Rate constant & Value & Units \\
 \hline\hline
 $k_1^a$ & 54 & $\rm M^{-1} s^{-1}$  \\ 
 \hline
 $k_{-1}^a$ & $8.6 \times 10^{-5}$ & $\rm s^{-1}$ \\
 \hline
 $k_1^c$ & 0.81 & $\rm \rm M^{-1} s^{-1}$\\
 \hline
 $k_{-1}^c$ & $1.29\times 10^{-6}$ & $\rm s^{-1}$ \\
 \hline
 $k_3^a$ & 0.01 & $\rm \rm M^{-1} s^{-1}$  \\ 
 \hline
 $k_{-3}^a$ & $5.8\times 10^{-8}$ & $\rm s^{-1}$\\ 
 \hline
 $k_3^c$ & 0.81 & $\rm \rm M^{-1} s^{-1}$  \\ 
 \hline
 $k_{-3}^c$ & $4.7\times 10^{-6}$ & $\rm s^{-1}$ \\ 
 \hline
 $k_{2,hv}$ & $1.8\times 10^{-3}$ & $\rm s^{-1}$  \\ 
 \hline
 $k_{-2,hv}$ & $0.5\times 10^{-6}$ & $\rm s^{-1}$  \\ 
 \hline
 $k_{2,\Delta}$ & $\approx10^{-16}$ & $\rm s^{-1}$  \\ 
 \hline
 $k_{-2,\Delta}$ & $1.3\times 10^{-6}$ & $\rm s^{-1}$  \\ 
 \hline
 $k_{4,hv}$ & $1.6\times 10^{-3}$ & $\rm s^{-1}$ \\ 
 \hline
 $k_{-4,hv}$ & $7.86\times 10^{-5}$ & $\rm s^{-1}$  \\ 
 \hline
 $k_{4,\Delta}$ & $\approx10^{-15}$ & $\rm s^{-1}$ \\ 
 \hline
 $k_{-4,\Delta}$ & $1.4\times 10^{-6}$ & $\rm s^{-1}$ \\ 
 \hline
 \end{tabular}
  \end{ruledtabular}
\end{table}

\section{Cycle statistics of tagged rings and axles}\label{sec_stat}

\subsection{Jump dynamics of rings and axles}
Let us imagine to follow a tagged ring or a tagged axle among the many other identical molecules. We would see that molecule jumping from one site to another; the ring can be found in one of 3 sites (free or attached to an axle in the {\em E} or {\em Z} form) and the axle in 4 sites (free in {\em E} or {\em Z} form, complexed with a ring in {\em E} or {\em Z} form). The jump from one site to another occurs when a reaction involving the tagged molecule takes place.

\begin{figure}[ht]
    \centering
    \includegraphics[width=0.45\textwidth]{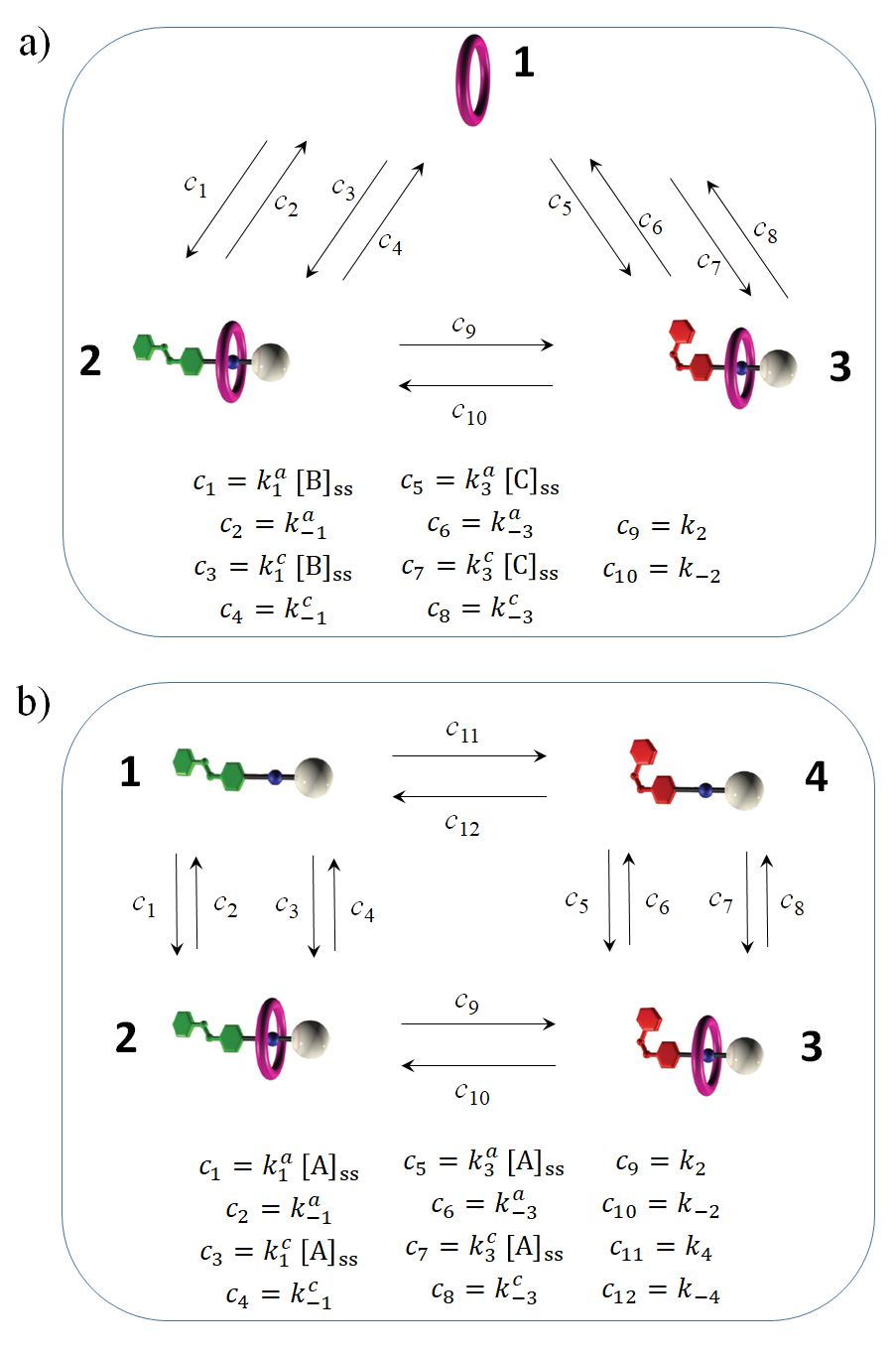}
    \caption{Jump processes of tagged molecules in the macroscopic context at the steady state. Panel a) shows the 3-sites scheme for a tagged ring; panel b) shows the 4-sites scheme for a tagged axle. For $[{\rm ring}]_{\rm tot} = 50 \, {\rm \mu M}$ and $[{\rm axle}]_{\rm tot} = 150 \, {\rm \mu M}$ (values used in the present simulations) and the rate constants in Table \ref{table1}, the steady-state concentrations are
$[{\rm A}]_{\rm ss} = 0.734 \, {\rm \mu M}$,
$[{\rm B}]_{\rm ss} = 4.69 \, {\rm \mu M}$
$[{\rm C}]_{\rm ss} = 96.0 \, {\rm \mu M}$
$[{\rm D}]_{\rm ss} = 0.147 \, {\rm \mu M}$
$[{\rm E}]_{\rm ss} = 49.2 \, {\rm \mu M}$. For the ring's dynamics,
$c_1 = 2.53 \times 10^{-4} \, {{\rm s}^{-1}}$,
$c_2 = 8.60 \times 10^{-5} \, {{\rm s}^{-1}}$,
$c_3 = 3.80 \times 10^{-6} \, {{\rm s}^{-1}}$,
$c_4 = 1.29 \times 10^{-6} \, {{\rm s}^{-1}}$,
$c_5 = 9.60 \times 10^{-7} \, {{\rm s}^{-1}}$,
$c_6 = 5.80 \times 10^{-8} \, {{\rm s}^{-1}}$,
$c_7 = 7.78 \times 10^{-5} \, {{\rm s}^{-1}}$,
$c_8 = 4.70 \times 10^{-6} \, {{\rm s}^{-1}}$,
$c_9 = 1.80 \times 10^{-3} \, {{\rm s}^{-1}}$,
$c_{10} = 1.80 \times 10^{-6} \, {{\rm s}^{-1}}$. 
For the axles,
$c_1 = 3.96 \times 10^{-5} \, {{\rm s}^{-1}}$,
$c_2 = 8.60 \times 10^{-5} \, {{\rm s}^{-1}}$,
$c_3 = 5.94 \times 10^{-7} \, {{\rm s}^{-1}}$,
$c_4 = 1.29 \times 10^{-6} \, {{\rm s}^{-1}}$,
$c_5 = 7.34 \times 10^{-9} \, {{\rm s}^{-1}}$,
$c_6 = 5.80 \times 10^{-8} \, {{\rm s}^{-1}}$,
$c_7 = 5.94 \times 10^{-7} \, {{\rm s}^{-1}}$,
$c_8 = 4.70 \times 10^{-6} \, {{\rm s}^{-1}}$,
$c_9 = 1.80 \times 10^{-3} \, {{\rm s}^{-1}}$,
$c_{10} = 1.80 \times 10^{-6} \, {{\rm s}^{-1}}$,
$c_{11} = 1.60 \times 10^{-3} \, {{\rm s}^{-1}}$,
$c_{12} = 8.00 \times 10^{-5} \, {{\rm s}^{-1}}$.    
    }
    \label{fig2}
\end{figure}

In the macroscopic limit, a tagged molecule is subjected to a Markov jump process among the available sites. Figure \ref{fig2} shows the reduced schemes that, for a tagged ring and for a tagged axle, can be derived from the pool of reactions according to what discussed in Sec. \ref{sec_macro}. 
 For the purposes of the present analysis, thermal and photochemical reactions have been merged into single effective channels: accordingly, 
$k_2 = k_{2, \Delta} + k_{2, h\nu}$,
$k_{-2} = k_{-2, \Delta} + k_{-2, h\nu}$
$k_4 = k_{4, \Delta} + k_{4, h\nu}$,
$k_{-4} = k_{-4, \Delta} + k_{-4, h\nu}$
(for details on the derivation of photochemical rate constants see Refs.
\cite{mauser1998photokinetics, penocchio2021photo}).
At the steady state, the site-to-site jump rate constants $c_{s \stackrel{m}{\to} s'}$, which for convenience have been labeled by a progressive number,
are fixed and bear a concentration dependence; the explicit expressions are given in Fig. \ref{fig2}.

In a small-volume closed environment of volume $V$, the system's state is specified by the array ${\bf n}(t) = \left[ 
n_{\rm A}(t) \; , \; 
n_{\rm B}(t) \; , \;
n_{\rm C}(t) \; , \;
n_{\rm D}(t) \; , \;
n_{\rm E}(t)
\right]$
 Given an initial array ${\bf n}_0 = {\bf n}(0)$, the paths in the space of the copy numbers are simulated by means of Gillespie's algorithm. In parallel with the global evolution of the whole system, we need to simulate the path of the tagged molecule through the available sites as described in Sec. \ref{sec_micro}.

\subsection{Direct and inverse cycles of tagged rings and axles}\label{sec_cycles}

Let us now introduce our definitions of {\em direct} and {\em inverse} cycles of a tagged ring and of a tagged axle. The corresponding cycle times will be indicated by $\tau$.

First, let us consider the cycles of the ring. Let the ring be initially in the free form (it is supposed that the ring has just completed its previous cycle by leaving the complex with an axle). Such a ring, sooner or later, will thread {\em some} available axle by entering from the side \textbf{a} (the axle, in principle, can be either in {\em E} or {\em Z} form, although the entrance in the {\em E} conformation is much easier). From this state, the ring can exit from the side \textbf{c} of the axle (again, the axle can be either in {\em E} or {\em Z} form, although the exit is much easier in the {\em Z} conformation). If this happens, we say that the ring has completed a {\em direct} cycle. Note that, while the ring is bound to the axle, several {\em cis-trans} transitions can take place. Also, the ring could exit from the side \textbf{a} too, so we would have to wait for the subsequent encounter with some available axle for observing the completion of the cycle. 
In a similar way, we define the {\em inverse} cycle of the ring just by inverting the sequence of events: the ring must thread some axle by entering from the side \textbf{c}, and then leave that axle by exiting from the side \textbf{a}.

Let us now consider the viewpoint of a tagged axle. That axle is initially in the free form (it is supposed that the previous cycle was just completed). Such an axle, sooner or later, will tether to {\em some} available ring entering from the side \textbf{a}. The axle, in principle, can be either in the {\em E} or {\em Z} form (although it is much more probable that the insertion of a ring occurs when the conformation of the axle is {\em E}). From such a state, the ring can exit from the side \textbf{c} of the axle (in the {\em E} or {\em Z} form). When this happens, we say that the tagged axle has completed a {\em direct} cycle. In between, the axle could undergo several {\em cis-trans} transitions. In a similar way, we define the inverse cycle: the axle has to tether to some ring entering from the side \textbf{c}, and then that ring must exit from the side \textbf{a}. 

 Of course, other definitions of cycles might be preferred. Note that the starting and ending points of the cycles correspond to the tagged molecules in their free state. While `free ring' corresponds to a unique site, the free axle may be either in the {\em E} or {\em Z} form. One might prefer to be more specific by requiring that the start/end points of tagged axle's cycle be the axle exactly in one of the two free forms, {\em E} or {\em Z}. The definitions of cycles given above are the same proposed in ref. \cite{sabatino2019b}, so to establish a continuation with the previous work.

Due to the aleatory character of the sequence of events and of the times at which each event takes place, the cycle time $\tau$ is a stochastic variable with its own characteristic statistical distribution. In the absence of irradiation, i.e. at thermal equilibrium, the statistics of the direct and inverse cycles are equivalent.
\footnote{
The ring would easily enter from the side \textbf{a} of an axle (likely in the {\em E} conformation), but it can also exit very easily from the same side while the required exit from the side \textbf{c} is more rare. In the opposite way, the entrance from the side \textbf{c} would be difficult but, after that, the inverse cycle would be easily completed by the quick exit from the side \textbf{a}. As a whole, the direct and the inverse cycles of the ring {\em must} have equivalent statistics at equilibrium.} 
The same happens for the cycles of the axle. 
On the contrary, the photoisomerizations break such equivalence by making the direct cycles quicker, on average, than the inverse ones. In fact, suppose that the ring enters from the side \textbf{a} of an axle (much probably in the {\em E} conformation). If a {\em trans}-to-{\em cis} isomerization takes place, likely a photo-induced one, exiting back from the side \textbf{a} is hindered, and hence the exit from \textbf{c} becomes the most feasible route. On the contrary, if the ring enters from the side \textbf{c} of an axle, the fact that under irradiation the axles are much probably in the {\em Z} conformation makes the exit from the side \textbf{a} more difficult to observe. A similar reasoning can be done for the cycles of the axle. As a whole, on average, the irradiation speeds up the direct cycles and slows down the inverse ones.

We note that, in spite of the fact that the cycles of rings and axles are different dynamical outputs with independent statistical distributions, it is {\em a priori} expected that the ratios $\overline{\tau}_{\rm axle, dir}/\overline{\tau}_{\rm ring, dir}$ (for the direct cycles) and
$\overline{\tau}_{\rm axle, inv}/\overline{\tau}_{\rm ring, inv}$ (for the inverse cycles)
are equal to the ratio between number of axles and number of rings in the sample (hence to the ratio between volumetric concentrations in the macroscopic setup). This can be rationalized by considering that the tagged ring eventually `engages' one single axle for completing its cycle and, similarly, the tagged axle eventually `engages' one single ring. As a consequence, every time a ring completes a direct (inverse) cycle, an axle also completes one, and vice-versa. By observing the whole sample in a certain time-window $\Delta t$, the number of direct cycles of rings and axles are therefore equal, and the same holds for the inverse cycles. Now imagine to observe the evolution of many independent replicas of the system for a very long $\Delta t$.
The average number of direct cycles of a single tagged ring is $\Delta t / \overline{\tau}_{\rm ring, dir}$. By considering that all rings are statistically equivalent and treated as independent under the tagged-molecule viewpoint (i.e., the possible dynamic correlations are implicitly included in $\overline{\tau}_{\rm ring, dir}$), we can express the average number of rings' direct cycles in the sample as $N_{\rm rings} \, \Delta t / \overline{\tau}_{\rm ring, dir}$. For the axles, the analogous relation holds: $N_{\rm axles} \, \Delta t / \overline{\tau}_{\rm axle, dir}$. By enforcing the equality between the two quantities, it follows that $\overline{\tau}_{\rm axle, dir} : \overline{\tau}_{\rm ring, dir} = N_{\rm axles} : N_{\rm rings}$. The same kind of relation is obtained for the inverse cycles.

\section{Numerical inspections}\label{sec_numerical}

\subsection{Physical setup}\label{sec_setup}
In order to follow the transition from macroscopic limit to small volume, the conditions in the small-volume situation have been fixed as follows. 
To make the comparison with the macroscopic case, we opt to choose numbers of rings and axles that exactly respect the macroscopic ratio rings:axles. Given the molar concentrations $[{\rm ring}]_{\rm tot}$ and  $[{\rm axle}]_{\rm tot}$, the volume of the compartment is fixed by  $V = n_{\rm rings} \, [{\rm ring}]_{\rm tot}^{-1} \, N_{\rm AV}^{-1}$ (or, equivalently, by $V = n_{\rm axles} \, [{\rm axle}]_{\rm tot}^{-1} \, N_{\rm AV}^{-1}$), where $N_{\rm AV}$ is the Avogadro number. 
In practice this would correspond to suddenly enclose a portion of solution of volume $V$, and isolate it from the rest of the sample.  
The shape of such a portion is irrelevant. Just for a physical visualization, we shall refer to a spherical shape thinking to vesicles or coacervates.
Of course, by imagining to repeat the compartmentalization experiment many times, the numbers of molecules would never be the same, and thus the fulfillment of the macroscopic ratio has to be intended on average. The choice we made, i.e. taking rings:axles equal to $[{\rm ring}]_{\rm tot} : [{\rm axle}]_{\rm tot}$, is only the most plausible reference situation.
\footnote{It is also true that lipid liposomes are nowadays easily synthesizable and this example might have some appeal. There is however a subtle issue with such a setup: for having the possibility of observing some dynamical effect of the compartmentalization, the radius of the vesicles should be of the order of the wavelengths of the radiation used to promote the photochemical reactions; this would imply inevitable scattering phenomena which complicate both the theoretical analysis and the interpretation of the experiments. Since here we are only investigating on the transition from macroscopic limit to small-volume sample, we ignore such a complication.}      

\begin{figure}[ht]
    \centering
    \includegraphics[width=0.45\textwidth]{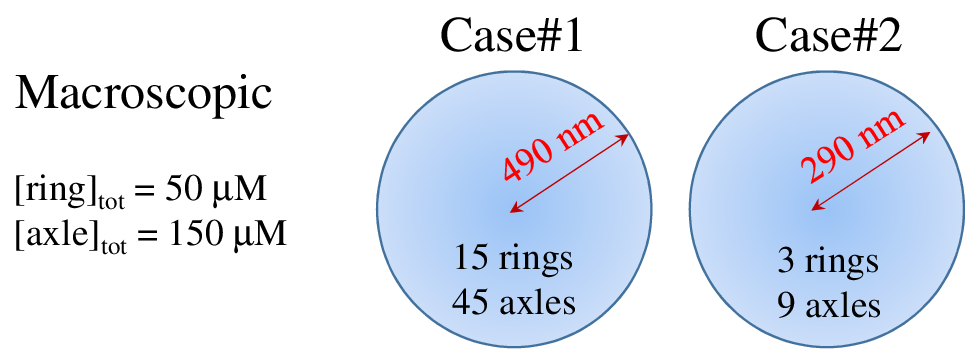}
    \caption{Pictorial representation of the physical setup: macroscopic reaction environment and small-volume cases.}
    \label{fig3}
\end{figure}

The simulations have been done for total concentrations $[{\rm ring}]_{\rm tot} = 50 \, {\rm \mu M}$ and $[{\rm axle}]_{\rm tot} = 150 \, {\rm \mu M}$, the same as in the experimental work of ref. \cite{ragazzon2015}. The steady-state concentrations of all species, and the values of the site-to-site jump rate constants in the macroscopic limit, are reported in the caption of Fig. \ref{fig2}.
In the small-volume context, the value of $V$ has been taken by imagining, as reactors, small spherical vesicles with radii of a few hundreds of nanometers. Two cases have been considered. In Case\#1, the simulations were run with 15 rings and 45 axles. The radius of the spherical compartment is 490 nm, corresponding to $V = 4.93 \times 10^{-19}$ L. In Case\#2, we consider 3 rings and 9 axles in a compartment of radius 290 nm and volume of $1.02 \times 10^{-19}$ L. 
Figure \ref{fig3} gives a pictorial representation of the compared situations. Some more inspections have been done for variants of Case\#1 and Case\#2 in which, at the same volume, the numbers of rings or axles were changed by one unit. The aim was to see if such little changes have some impact on the cycles' timing when the total number of molecules is very small.

\subsection{Computational details}\label{sec_computdet}

Fortran codes have been written for the simulations and the post-production analysis. All random number generations in the interval $[0,1]$ have been done by means of the subroutine `ran2' (from Numerical Recipes\cite{NR}) initialized with the computer's clock. In the macroscopic context, given $[{\rm ring}]_{\rm tot}$, $[{\rm axle}]_{\rm tot}$, and the rate constants in Table \ref{table1}, the steady-state concentrations have been numerically determined by means of the implicit solver for stiff kinetics `Variable-coefficient ODE solver' (VODE) \cite{VODE} up to reach a time sufficiently long to ensure stable values (tolerated variations at most of 1\%).
\footnote{
The computations have been done with the Fortran77 double-precision subroutine `DVODE' freely available at
https://computing.llnl.gov/projects/odepack/software. Last viewed on 17 January 2022.}

The simulations of tagged molecule paths have been done by means of Gillespie's SSA as described in sections \ref{sec_macro} and \ref{sec_micro}. 
The initial sites were set to be free ring for the tagged ring, and free {\em E}-axle for the tagged axle (such choices are however immaterial). The completion of the direct and inverse cycles was detected on the fly until the required number of cycles was achieved. Independent simulations have been done for all four instances of direct/inverse cycle of ring/axle.

\begin{figure}[ht]
    \centering
    \includegraphics[width=0.45\textwidth]{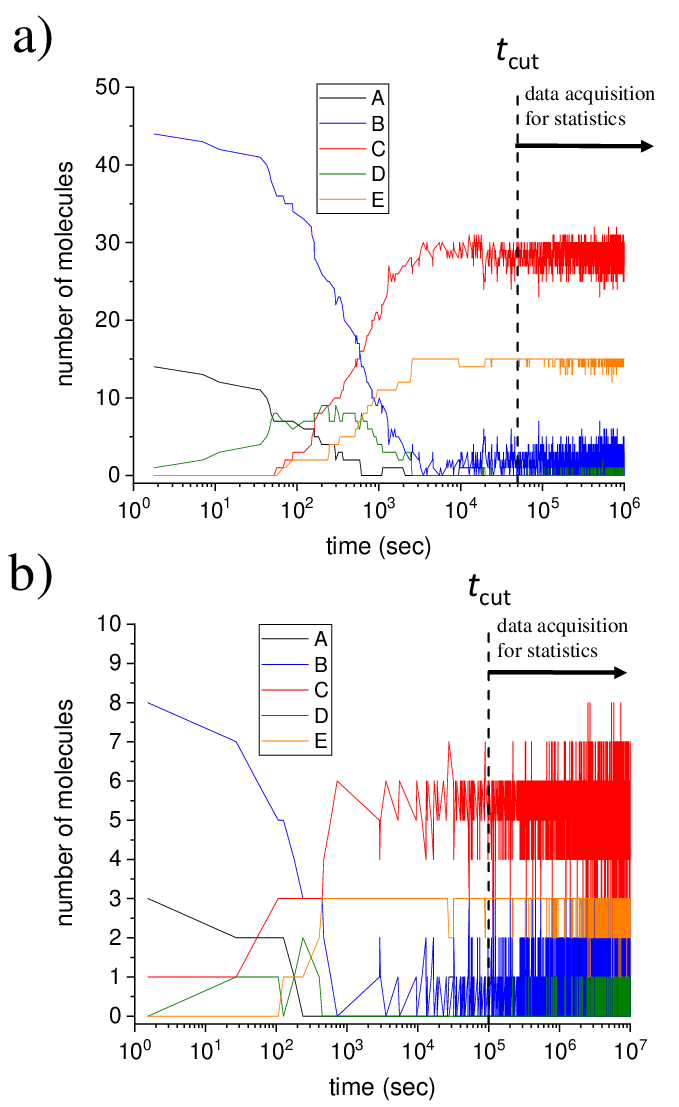}
    \caption{Examples of temporal evolution of the molecule copy numbers in the small-volume Case\#1 (panel a) and Case\#2 (panel b).}
    \label{fig4}
\end{figure}

In the small-volume cases, the duration of the initial transient phase to be excluded in the statistical analysis was directly assessed by looking at the temporal profiles of the numbers of molecules of each species. When the fluctuations do stabilize around average values for each species, the transient phase was considered to be concluded. To be safe, a longer time ($t_{\rm cut}$) was taken before acquiring the data for the statistics. Examples of temporal profiles for Case\#1 and Case\#2 are given in Fig. \ref{fig4} where the applied $t_{\rm cut}$ is indicated. It was checked that the statistical outcomes are reproducible taking a smaller $t_{\rm cut}$, so ensuring that the steady state conditions were indeed reached.
In the simulations, as initial condition it was set that all rings are in the free form (species A) and all axes free in the {\em E} conformation (species B).

For the statistics, in the macroscopic limit the number of cycles (direct or inverse) was $10^7$  for both ring and axle. In the small-volume context, the much heavier computations required to lower the number of cycles. The cycles were still $10^7$ for the direct cycles of ring and axle, and $10^6$ for the inverse cycles. In all cases the quality of the statistics was assessed by checking that the average cycle time and the standard deviation of the data (later shown in Table \ref{table2}) were essentially stable under lowering the number of cycles by a factor 10. The variations were within 1\% in all cases. 
Although a statistical uncertainty on the outcomes cannot be provided because of the difficulty of repeating the simulations for a sufficiently large number of times, such   1\% variation is an indicator of both convergence and reproducibility of the outcomes. In few repetitions of the simulations made as check, the results were indeed within such a range of variation. 
The distributions of the cycle times have been obtained by histograms with a logarithmic binning to magnify the details at low values of the cycle times; the number of bins was 200 for the direct cycles, and only 20 for the inverse cycles because of the lower number of data. Despite the limited quality of statistics for the inverse cycles, in all cases the features of the distributions (in particular the rising of the left tails) are enough well-defined for our discussion.

\subsection{Results and discussion}

\begin{figure*}[ht]
    \centering
    \includegraphics[width=0.9\textwidth]{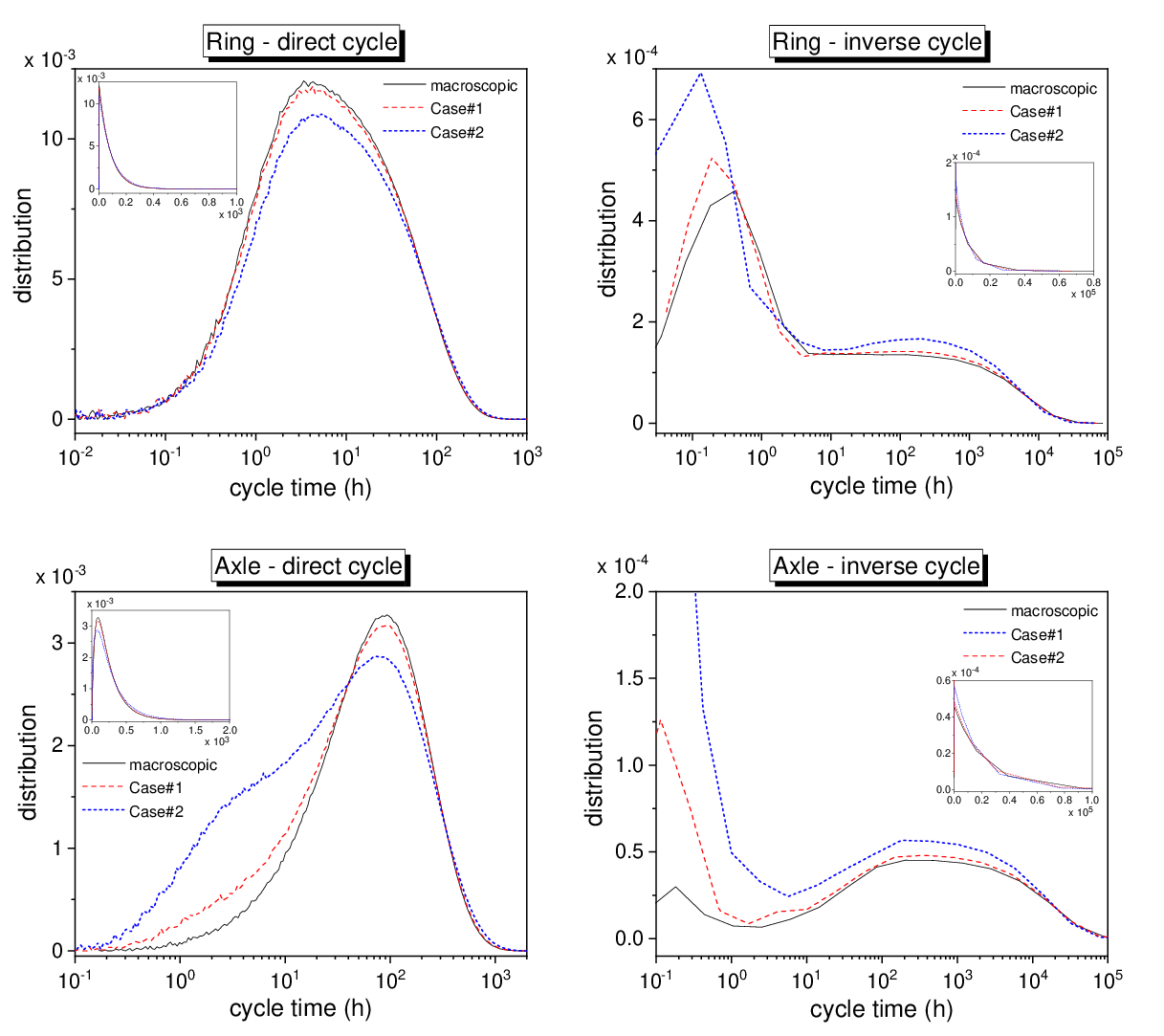}
    \caption{Statistical distributions of the cycle times of tagged rings and axles. The continuous black lines refer to the macroscopic context, the dashed red lines refer to the small volume Case\#1, and the short-dash blue lines to Case\#2. Note the logarithmic scale on the time axis. The insets show the same profiles on linear scale.}
    \label{fig5}
\end{figure*}

 The distributions of the cycle times for the direct and inverse cycles of tagged rings and axles are shown in Fig. \ref{fig5}. 
 The black lines refer to the macroscopic limit,
 the dashed red lines to the small-volume Case\#1, and the short-dash blue lines to Case\#2. The profiles for the macroscopic case are comparable to those presented in ref. \cite{sabatino2019b}. 
Table \ref{table2} collects, for all instances, the average cycle time and the standard deviation of the distribution as obtained from the data.
 
What appears from the profiles in Fig. \ref{fig5} is that, in passing from macroscopic sample to small volume, the distributions differ only in the left tail at short cycle times. However, this corresponds to a narrow portion of the whole distribution, which is magnified by the representation in the logarithmic scale. In the linear scale, as shown in the insets, such a portion is hardly detectable and, in practice, it gives a little or even negligible contribution to the average values in Table \ref{table2}. The ring's direct cycle appears to be the less affected by the transition from macroscopic sample to small volume. In the other cases, the reduction of the sample volume seems to raise the left tails and hence to promote some paths that allow to complete the cycles quickly.

Concerning the average cycle times and the standard deviations, some preliminary comments are in order. 
First, for both direct and inverse cycles, the ratio between the average cycle times of axle and ring is equal to 3 (apart from negligible deviations due to the finiteness of the statistical ensemble of simulations), which corresponds to the ratio between number of axles and number of rings. This is in agreement with the {\em a priori} expectation as discussed at the end of Sec. \ref{sec_cycles}; the nearly perfect matching constitutes an indirect check of the good quality of the statistics, especially considering that each instance was simulated independently from the others.
Second, the standard deviations are very close to the average values. This is because the raising left tail of the distributions plays little role as said above and, in the linear scale, the profiles look very close to mono-exponential decaying distributions for which average and standard deviation do coincide.
This implies that the so-called squared coefficient of variation (i.e., the square of the ratio between standard deviation and average) is nearly equal to 1. Such a quantity, also known in statistical kinetics as `randomness parameter'\cite{moffit2014, Fisher07, Block95}, is particularly relevant since in many cases it can be experimentally determined thanks to single-molecule techniques nowadays available.

A global look at the values in Table \ref{table2} makes us stating that, in all cases, the values for macroscopic and small-volume situations are comparable. 
Even for Case\#2, despite the numbers of involved molecules are very small, there is little difference with respect to the macroscopic limit. From the data in Table \ref{table2}, it emerges that in going to such small numbers of molecules the direct cycles are on average slightly slowed down, while the inverse cycles are on average sped up. 
This is actually an empirical result from the present simulations done with the rate constants of Table \ref{fig1}. To have appreciable differences one has to deal with very small numbers of rings and axles, in the order of a few units. Even in a compartment of about 300 nm of radius containing only 3 rings and 9 axles, however, the average times of the inverse cycles decrease only by about 20\% with respect to the macroscopic limit.

\begin{table}
    \caption{Average cycle time, $\overline{\tau}$, and standard deviation of the data for the direct and inverse cycles of rings and axles in the macroscopic limit and in the small-volume situations of Case\#1 and Case\#2. Times and standard deviations are expressed in hours. The numbers of data used for the statistics are given in Sec. \ref{sec_computdet}. The uncertainty on the data reported in the table is of the order of 1\%.}
    \label{table2}
\begin{ruledtabular}
\begin{tabular}{c cc  cc cc}
\multicolumn{7}{c}{DIRECT CYCLES} \\
&\multicolumn{2}{c}{macroscopic} &\multicolumn{2}{c}{Case\#1} &\multicolumn{2}{c}{Case\#2} \\
& $\overline{\tau}$ & St. Dev. & $\overline{\tau}$ & St. Dev. & $\overline{\tau}$ & St. Dev. \\  
RING & 80.2 & 79.3 & 81.9 & 81.0 & 88.7 & 87.6 \\
AXLE & 240.7 & 198.2 & 245.9 & 205.7 & 
266.2 & 235.8 \\
\hline
\multicolumn{7}{c}{INVERSE CYCLES} \\
&\multicolumn{2}{c}{macroscopic} &\multicolumn{2}{c}{Case\#1} &\multicolumn{2}{c}{Case\#2} \\
& $\overline{\tau}$ & St. Dev. & $\overline{\tau}$ & St. Dev. & $\overline{\tau}$ & St. Dev. \\  
RING & 7486 & 7490 & 7082 & 7092 & 5977 & 5971 \\
AXLE & 22458 & 22423 & 21285 & 21225 & 
17919 & 17903 \\
 \end{tabular}
 \end{ruledtabular}
\end{table}

The loose dependence of the average cycle times on the sample size might seem surprising. Despite the fluctuations of the copy numbers are marked (see Fig. \ref{fig4}), the average times are almost insensitive to such fluctuations. This is likely due to the fact that the bimolecular reactions are not the rate-limiting processes even when the numbers of molecules become very small.

Let us look in more detail to such an aspect for the ring-axle scheme
with the adopted parameters. Focusing on the direct cycles of the ring, the fastest process of cycle initiation is the entrance of the ring from the side {\bf a} of an {\em E}-axle. The jump rate associated with such bimolecular process is $c_1 = 2.53 \times 10^{-4} \, {\rm s^{-1}}$, 
which is higher than the rates of the unimolecular processes (except the {\em E} to {\em Z} transition of the complexed axle with rate constant $k_2$). Hence,
in the macroscopic limit the bimolecular processes are not rate-limiting. The same is true in the small-volume context. In fact, the rate factors (see Eq. \ref{fprop}) entering the propensity functions of the bimolecular reactions of initiation are $f_1^a = 1.8 \times 10^{-4} \, {\rm s^{-1}}$ and
$f_3^a = 3.4 \times 10^{-8} \, {\rm s^{-1}}$ for Case\#1, and $f_1^a = 8.8 \times 10^{-4} \, {\rm s^{-1}}$ and
$f_3^a = 1.6 \times 10^{-7} \, {\rm s^{-1}}$ for Case\#2. Roughly speaking, the associated jump rate constants (for the jumps from free ring to bound ring) are of the order of such factors multiplied by the typical number of {\em E}-axles or {\em Z}-axles under steady conditions. This tells us that also in the small-volume context the entrance of the ring from the side {\bf a} of an {\em E}-axle (the main initiating process) is not rate-limiting. In short, the unimolecular processes remain essentially the rate-limiting ones at all scales. For the inverse cycle, the initiating bimolecular processes are associated with $c_3 = 3.8 \times 10^{-6} \, {\rm s^{-1}}$ (entrance through the side {\bf c} of a {\em E}-axle) and
$c_7 = 7.8 \times 10^{-5} \, {\rm s^{-1}}$ (entrance through the side {\bf c} of a {\em Z}-axle).
The main initiating process is the entrance through the side {\bf c}, whose rate constant is comparable to, or higher than, the rates of the unimolecular processes of exit from the side {\bf a}.
Thus, even in this case, in the macroscopic sample the bimolecular processes are not the rate-limiting ones. This holds true also in the small-volume context since $f_1^c = f_3^c = 2.7 \times 10^{-6} \, {\rm s^{-1}}$ for Case\#1 and $f_1^c = f_3^c = 1.3 \times 10^{-5} \, {\rm s^{-1}}$ for Case\#2 are by themselves comparable with $c_3$ and $c_7$.
As a whole, even in the limit of very few numbers of molecules, the timing of the inverse cycle of a ring should not markedly differ from the macroscopic situation, as it is indeed observed. A similar reasoning can be done for the cycles of the axle. A borderline situation is encountered for the axle's inverse cycle, for which it is found that the bimolecular processes are associated with jump rate constants comparable to those of the unimolecular processes.

To summarize, the simulations for the ring-axle system support the idea that the compartmentalization affects mainly the left tail of the distributions of the cycle times (promoting the paths that allow the quick completion of the cycles), but has little effect on the average behaviour (average cycle times and standard deviations). 
At first sight, the effect on the narrow left tails of the cycle time distributions might seem a minor feature. On the other hand, suppose that completing the cycles within a certain threshold time is relevant for some reasons (we are just reasoning in abstract terms), then the narrow left tail would become the main feature.  
Concerning the weak dependence of the average behaviour on the sample volume, it is likely due to the fact that the rate-limiting processes are some of the unimolecular reactions, at least under the conditions here explored. Changing the conditions (reaction rate constants and total concentrations of rings and axles) the outcomes could be different case by case, but the above line of reasoning remains valid.

To conclude this section, let us briefly comment on the effects of increasing by only one unit the number of rings or axles in Case\#1 and Case\#2, while keeping the volume fixed. We performed such additional computations to highlight how little variations in the number of molecules within a small compartment may affect the dynamical output of tagged molecules. In fact, such kind of variations are expected in a real compartmentalization experiment, as the replicas will much probably differ one from each other, thus introducing a sort of “static disorder”. This is for instance analogous to what happens in stochastic gene expression, where it addressed as ‘extrinsic noise’\cite{Elowitz02, Raser05}. A similar kind of static disorder has been inspected at the individual-molecule level also in catalytic processes with catalysts supported on amorphous materials \cite{Skodje21}. Here, by changing the number of axles in Case\#2 from 9 to 10  (keeping the number of rings equal to 3) we did not observe significant variations in the average cycle times of the rings, while the average cycle times of the axles of both direct and inverse cycles do increase by about 10\%. This is actually the expected trend, since the tagged axle has to ‘compete’ with more other axles to engage one available ring and complete a cycle. When the number of rings changes from 3 to 4 (keeping the number of axles equal to 9), the average cycle times of the rings still remain nearly unaltered, while the average cycle times of the axles decrease by about 25\%. This is because, in such a case, more rings are available and the completion of the cycles by the tagged axle is therefore facilitated. Similar yet less pronounced trends are found for the analogous variants of Case\#1, with variations of about 2\% when passing from 45 to 46 axles, and of about 6\%  when passing from 15 to 16 rings.

\section{Remarks and conclusions}\label{sec_results}

In this work, we have presented the general approach for inspecting dynamical aspects of steady-state reaction networks from the viewpoint of individual tagged molecules that change their chemical state upon occurrence of reactions (we recall that the approach can be even extended to individual molecular moieties\cite{sabatino2019a}). The specific aspect investigated here is the possible dependence of an individual-molecule dynamical output on the size of the sample, going from the macroscopic limit to small-volume situations in which the kinetics are inherently stochastic due to the small numbers of reactive molecules. In making such a change of scale, we think to a sort of compartmentalization in which a smaller and smaller portion of a well-mixed sample is isolated from the rest of the solution.

It has been pointed out that a size dependence can be ascribed to the presence of bimolecular reactions because the related propensity functions bear a volume dependence. Although this is intuitive, a quantitative assessment of the size dependence requires to simulate the tagged molecule path by using the tools described here. 
The method has been applied to a case study of relevance in the broad field of artificial molecular machines, namely a light-driven bimolecular motor composed by ring-like and axle-like molecules.
By adopting the same physical parameters of ref. \cite{sabatino2019b} and the same experimental conditions of ref. \cite{ragazzon2015}, it turned out that the principal descriptors of ring's and axle's cycle time statistics (average times and standard deviations) are weakly dependent on the size of the sample. However, the shape 
of the left side of the cycle time distributions changes when the numbers of molecules become small and the fluctuations are strong.

The weak sample size dependence of the average behaviour has been explained by pointing out that the bimolecular processes are never rate-limiting at all the scales. Of course, we cannot exclude that other statistical features (i.e., other dynamical outputs) of the ring-axle dynamics are more sensitive to the sample size when approaching the limit of very small numbers of molecules. The present analysis is therefore by no means exhaustive, and addressing other aspects would require further investigations.

The outcomes for the present system make us claiming that the robustness (on average) of a tagged-molecule dynamical output 
to the compartmentalization in small volumes
might be ascribed to the presence of unimolecular reactions in the network, and that such processes must be rate-limiting at all scales (or, better, the bimolecular processes do not become rate-limiting). In this regard, it would be interesting to inspect various instances of natural biochemical networks to see how unimolecular and bimolecular reactions do concur to regulate the jump dynamics of the functional molecules. Another interesting aspect, to be further inspected, concerns the fact that the compartmentalization might selectively alter the statistical weights of the tagged-molecule paths. In the ring-axle case this is related to the observed raising of the left tails of the cycle time distributions, which is intuitively associated with the promotion of paths that allow a quick completion of the cycles.  
It would be interesting to inspect if the compartmentalization has similar effects on the dynamics of tagged molecules in other reaction networks. 

We conclude with a note of caution recalling that one basic assumption of our approach is that the compartmentalization does not affect the intrinsic kinetic parameters. The sample's delimitation here enters only through the volume as a scaling factor for the bimolecular propensity functions. In crowded microreactors, like the coacervates mentioned in the Introduction\cite{coac1, coac2}, additional case-dependent subtle effects, like the stabilization of transition states, might be relevant. In particular, such specific effects might have to be necessarily taken into account for explaining the kinetic response when only unimolecular reactions are present.


\begin{acknowledgments}
The research activity of D. A. has been carry out during an Erasmus+ Traineeship at the University of Luxembourg. We thank Massimiliano Esposito for having hosted D. A. in his research Group and for the useful opinions on the work.   
E.P. was funded by the European Research Council project NanoThermo (ERC-2015-CoG Agreement No. 681456).
\end{acknowledgments}

\section*{Data Availability Statement}
The data that support the findings of this study are available within the article.

\section*{Reprints and permissions}
This article may be downloaded for personal use only.
Any other use requires prior permission of the author and AIP Publishing.
This article appeared in \textit{The Journal of Chemical Physics} \textbf{156}, 184116 (2022) and may be found at https://doi.org/10.1063/5.0089695.

\end{document}